# A FORMAL ARCHITECTURE-CENTRIC MODEL-DRIVEN APPROACH FOR THE AUTOMATIC GENERATION OF GRID APPLICATIONS

David Manset[1,2,3]
2 ETT Division, CERN 1211, Geneva 23, Switzerland
david.manset@cern.ch

Hervé Verjus[3]
3 LISTIC, University of Savoie, Annecy, France
herve.verjus@univ-savoie.fr

Richard McClatchey[1]
1 CCCS, University West of England, Bristol, UK
richard.mcclatchey@cern.ch

Flavio Oquendo[4]
4 VALORIA, University of South Brittany, Vannes, France
Flavio.Oquendo@univ-ubs.fr

Abstract: This paper discusses the concept of model-driven software engineering applied to the Grid application domain. As an extension to this concept, the approach described here, attempts to combine both formal architecture-centric and model-driven paradigms. It is a commonly recognized statement that Grid systems have seldom been designed using formal techniques although from past experience such techniques have shown advantages. This paper advocates a formal engineering approach to Grid system developments in an effort to contribute to the rigorous development of Grids software architectures. This approach addresses quality of service and cross-platform developments by applying the model-driven paradigm to a formal architecture-centric engineering method. This combination benefits from a formal semantic description power in addition to model-based transformations. The result of such a novel combined concept promotes the re-use of design models and facilitates developments in Grid computing.

## 1. INTRODUCTION

The Grid paradigm is described in (Foster et al, 2001) as "a distributed computing infrastructure for advanced science and engineering" that can address the concept of "coordinated resource sharing and problem solving in dynamic, multi-institutional virtual organizations". This coordinated sharing may be not only file exchange but can also provide direct access to computers, software, data and other system resources. Grid applications bundle different services using a heterogeneous pool of resources in a so-called *virtual organization*. This makes Grid applications very difficult to model and to implement.

In addition, one of the major issues in today's Grid engineering is that it often follows a code-driven approach. Although it has been proven from past experience that using structured engineering methods would ease the development process of any computing system and would reduce complexity, the inter-disciplinarily of Grid computing is still encouraging 'brute-force' coding and consequently a rather unstructured engineering process. This always leads to a loss of performance, interoperability problems and generally ends in very complex systems that only dedicated and expert developers can manage. As a direct consequence the resulting source code is neither re-usable nor does it promote dynamic adaptation facilities as if it were a true representation of the Service Oriented Architecture (SOA). Having no guidelines or rules in the design of a Grid-based application is a paradox since there are many existing, architectural approaches for distributed computing which could ease the engineering process, could enable rigorous engineering and could promote the re-use (Cox, 2004) of software components in future Grid developments.

It is our belief that code-driven approaches and semi-formal engineering methods in current use are insufficient to tackle tomorrow's Grid developments. This paper provides a set of Grid specific models enacted within a novel engineering approach that implements the model-driven philosophy. Inside a well-defined and adapted formal approach, we investigate the enactment of our model-driven engineering process providing the tools to build the next generation of Grid applications. Thus, this paper emphasizes different aspects, which are, in our view, essential to Grid engineering:

- it offers a user-friendly vision to Grid architects by providing re-usable conceptual building blocks,
- it hides the complexity of the final execution platform through abstraction models,
- it promotes design re-use to facilitate further developments.

To achieve these objectives, we combine two approaches together and seek advantages from each.

On the one hand, we use the formal semantic descriptive power to model Grid applications; on the other hand, we use a model-driven approach to promote model re-use, model transformations, to hide the platform complexity and to refine abstract software descriptions to concrete ones.

The remainder of this paper is structured as follows. Part 2 presents the approaches used, (i.e. Model Driven Engineering and Architecture-centric approach). Part 3 explains how model-driven engineering is enacted to design Grid applications. Part 4 presents our formal architecture-centric model-driven approach and the means used to achieve it. Part 5 illustrates the presented paradigms with a concrete example. Finally, we conclude with identifying future work and state the benefits of using the presented concept.

## 2. MODEL-DRIVEN AND ARCHITECTURE CENTRIC APPROACHES

### 2.1. The MDE Approach

Model Driven Engineering MDE (Kent, 2002), probably derived from the OMG Model Driven Architecture MDA$^{TM}$ (Kleppe et al, 2003) initiative, tackles the problem of system development by promoting the usage of models as the primary artefact to be constructed and maintained.

Enacting the model-driven paradigm is not an easy route to follow since as yet there are few available frameworks designed and most of them are combinations of existing tools. Despite the lack of proper MDE tools, there are clear advantages from using and enacting it. Among such benefits that are valuable for Grid applications is providing system developers the capability to design systems efficiently in a heterogeneous and rapidly changing environment. Indeed, models being decoupled from platform technologies, system descriptions remain relevant and re-usable.

## 2.2. A Combination of Approaches

By convention and in order to separate clearly the concept described here from that of the OMG MDA[TM], we call our approach a *grid model-driven engineering* approach *(gMDE)* and use a Grid-specific terminology. The OMG describes a design method based on model transformations according to meta-models, which is generic enough to fulfil any requirements in terms of modelling and re-use. However, most existing implementations of this paradigm provide only model to source code transformations, based on UML, where the Platform Independent Models *(PIMs)* are translated to Platform Specific Models *(PSMs)*. In Grid engineering, when mapping system models to concrete platforms, it is often necessary to include model to model transformations to fill the gap between the abstract description and its concrete representation. In addition, model optimization requires the generation of intermediate models to compute and synchronize different views of a system. Providing model to model transformations as well as model to code transformations along the development process makes the approach more modular and also facilitates the final source code generation. In order to support this approach, we combine the model-driven philosophy to a well-established architecture-centric approach (Chaudet et al, 2000).

Thus, we first define a set of key models to design a Grid application from the high level descriptions of each architectural element to its final deployment. In addition, we introduce the necessary semantics to generate, transform and check models along the design process. We consider architectural descriptions (from abstract to more concrete) as models. From this basis, transformations are applied to models and as a consequence to software architectures according to architect's and platforms requirements. The end result of such iterative modifications and mappings being the concrete deployed application.

Focusing on the model transformation aspects, we can notice similarities with the refinement concepts found in formal architecture-centric software engineering developments. The MDE can benefit from refinement to handle some of the model transformations and to ensure the models' correctness. From the OMG's vision, we use the basic idea that consists of starting from a PIM to go to a PSM by means of transformations. Our approach follows such an idea in implementing the architecture-centric refinement (see section 3).

## 3. A FORMAL ARCHITECTURE-CENTRIC MDE APPROACH

Following our *gMDE* paradigm (Manset et al, 2005), we address the challenge of designing, optimizing and adapting Grid-abstract architectures, with respect to different criteria, in order to automatically generate a complete set of Grid services to be deployed on a physical grid of resources. From the work we conducted in Grid engineering (Amendolia et al, 2005) we consider the Grid as a SOA and provide the means to specify system properties related to the Quality of Services QoS (Land, 2002) and Grid middleware platforms. Using formal semantics, we build a set of major models and investigate their orchestration along the *gMDE* design process.

### 3.1. The gMDE Key Models

In Grid engineering, design is largely affected by many constraints; these constraints are of different types and are introduced either by the architect when implementing QoS related features or by the target execution platform. Thus the MDE process dedicated to Grid engineering must take into account all of these aspects in providing the necessary models and semantics. By proposing several models (see figure 1), our approach separates concerns and addresses different aspects of Grid applications. Thus expertise management and capture are better than in classical approaches e.g. (Medvidovic et al, 1996). Each model represents an accurate aspect of the system, useful for conceptual understanding, analysis and refinement. Unlike the software engineering process where the system architecture is iteratively refined by the architect, most of the transformations in the *gMDE* are automated. The different models composing our process are defined as follows:

- *GEIM – Grid Environment Independent Model*:
an abstract description of the Grid application based on a formal ADL (Architecture Description Language) – using domain specific constructs,

- *GESM – Grid Environment Specific Model*:
a concrete architecture close to the final code and optimized according to a particular Grid middleware (execution platform) and QoS properties (a refined system description),
- *GECM – Grid Execution Constraint Model*: a design pattern representing a particular QoS property,
- *GETM – Grid Environment Transformation Model*: a design pattern representing a particular Grid platform.

As a clarification of concept, we do not discuss in detail the other models composing our design process. However, these models can be defined as follows:
- *GEMM – Grid Environment Mapping Model*: a model of translation between an architecture description language and an implementation language (i.e. that defines the mapping between the semantics of the *GESM* and a given programming language, for instance Java).
- *GERM – Grid Environment Resource Model*: a model representing the physical constitution of the Grid.
- GEDM – Grid Environment Deployment Model: a model specifying the distribution and deployment of the resulting application onto the grid set of resources,
- *GESA – Grid Environment Specific Application*: the auto-generated source code of the application (i.e. obtained after *GEMM* translation).

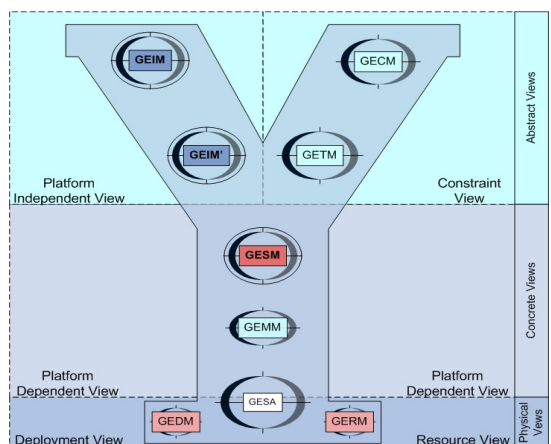

Figure 1. gMDE key models

Figure 1 expresses the progressive design convergence of these models towards the generation of the final system source code (*GESA*) and its deployment over the physical infrastructure. This convergence is punctuated by different transformations (in nature and objectives).

As is mentioned in section 2, our model-driven approach uses the architecture-centric refinement concept to decouple:
- the abstract domain specific vision from the concrete implementation and
- the architect's functional specifications and non-functional requirements.

As is depicted in figure 1, the models represent different views of the system. Typically, non-functional aspects – referred to as "Constraint View" - are defined inside the *GECM* and *GETM* models; unlike functional aspects – referred to as "Platform Independent View" - which are defined in the *GEIM* and *GEIM'* (a specialized form of the *GEIM*) models.

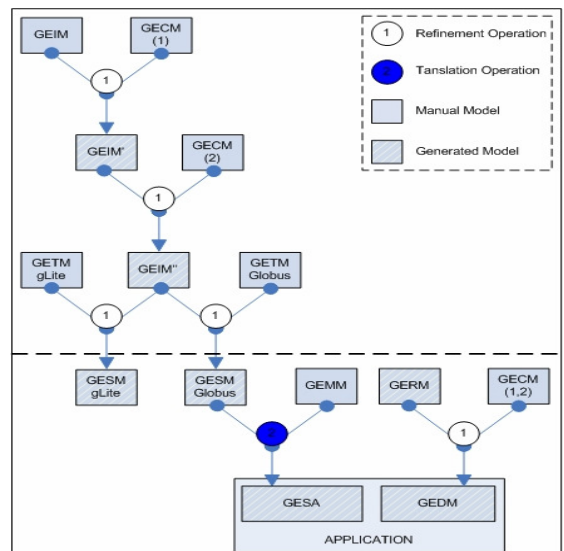

Figure 2. The gMDE development process

Each of these two views owns a proper meta-model introducing a Grid domain terminology to facilitate the domain representation. Once the concrete system specification (*GESM*) has been obtained, it is translated into source code (*GESA*) using a mapping expressed in the *GEMM* model. This transformation and corresponding models are referred to as part of the "Concrete View". Finally, the system distribution over a physical set of resources (an essential aspect of Grid computing) is also handled using models (the *GEDM* and the *GERM*) and transformations, constituting the "Physical View".

The GEIM, GETM and GESA are the only models visible and modifiable by the software architect during design, unlike others, which are automatically obtained from transformations.

## 3.2. The gMDE Development Process

Figure 2 introduces the orchestration of the previously presented models inside the *gMDE* design process. In the depicted process, a distinction is made between two major levels, one is the architecture level of transformation – above the broken lines - and the other is the implementation level of transformation – below the broken lines. Models and transformations can differ in nature and objectives. Thus models can be of two distinct types; either the model is manually created or it is automatically obtained by transformation. Transformations can then be of two different types; either the transformation is a composition of one or more refinement actions (model to model transformation) or it is a translation mapping (model to source code transformation).

In the previous drawing, two different sets of QoS constraints were successively introduced (referred to as *GECM 1 & 2*). By introducing new models, the software architect can specialize an architecture progressively with respect to different sets of constraints. Once the system architecture complies with the expressed requirements, the software architect can specify a Grid execution platform. This is illustrated in figure 2 (referred to as the *GETM* for (gLite) and for (Globus) *Grid platforms*), two different middlewares were selected to obtain the adapted concrete system architecture, *GESM*. Figure 2 also details the models and transformation types. The depicted process demonstrates the integration of multiple constraints by the introduction of models. The gMDE approach covers both model to model transformations and model to code transformations, which makes it flexible enough to tackle other aspects. Indeed, the process is not limited to what is expressed in figure 2 but can be extended to any sets of constraints, provided the corresponding model is expressed. This scalability is the direct result of the underlying formal architecture-centric model-driven approach.

# 4. ENACTING MDE, A CONCRETE FRAMEWORK

## 4.1. ArchWare: Formal Architecture-centric Approach and Toolkit

(ArchWare) is an engineering environment supporting the development of software systems through the use of a formal architecture-centric approach. This formal architecture-centric method enables the support of critical correctness requirements and provides tools to guarantee system properties. ArchWare provides a set of formal languages to enable reliable design, amongst them: (1) the ArchWare Architecture Description Language ADL (Oquendo et al, 2002), defined as a layered language for supporting both structural and behavioural descriptions as well as property definitions. This language is based on the π-calculus (Milner, 1999) and μ-calculus (Kozen, 1983), (2) the ArchWare Architecture Refinement Language ARL (Oquendo et al, 2004), used to describe software architectures (based on the Component and Connector architectural style) and to refine them accordingly to transformation rules.

These languages used together constitute the ArchWare environment framework. As mentioned in section 2, there are noticeable conceptual similarities between some of the *gMDE* model transformations and software architecture refinement operations. From our point of view, refinement is considered as an architecture-level transformation. Thus, the rest of this paper investigates the ArchWare refinement process ,which is, we believe, essential to the enactment of our formal architecture-centric MDE approach.

## 4.2. The ArchWare Refinement Concept

Complex systems cannot be designed in one single step. In a stepwise architecture refinement, a sequence of modifications is applied on a system abstract model, which leads to a concrete, implementation-centred model of the architecture. These refinement steps can be carried out along two directions: "vertical" and "horizontal". The concrete architecture of a large software system is often developed through a "vertical" hierarchy of related architectures. An architecture hierarchy is a linear sequence of two or more architectures that may differ with respect to a variety of aspects. In general, an abstract architecture is simpler and easier to understand, while a concrete architecture reflects more implementation concerns. "Vertical" refinement steps add more and more details to abstract models until the concrete architectural model has been described. A refinement step typically leads to a more detailed architectural model that increases the determinism while implying properties of the abstract model. "Horizontal" refinement concerns the application of different refinement actions on different parts of the same abstract architecture, for instance, by partitioning an abstract component into different pieces at the same abstraction level. The ArchWare ARL language is the formal expression of these refinement operations, which aims at preserving upper abstract architecture properties while modifying it. The ArchWare environment supports at each level of the design process the re-use of existing architectural models and, at the concrete level, architecture-based code generation. As is demonstrated in [24], the ArchWare approach handles an exhaustive set of refinement actions. The semantics of such actions are expressed as follows:

```
refDefinition::=on a : architecture action actionName is refinement (
actionParameter_0 , actionParameter_n )
{
            [ pre is { condition } ]
            [ post is { condition } ]
            [ transformation is { refExpression } ]
} [ assuming { property } ]
```

Each refinement action, hereinbefore referred to as *actionName*, specifies a refinement action to apply on an architecture "a", as well as pre- and post-conditions.

### 4.3. A Refinement Process for gMDE

The *gMDE* approach focuses on both directions of refinement i.e. the "vertical" and the "horizontal". The intention is not only to refine an architecture to a concrete and "close to final" code form, but also to adapt it according to constraints. This paper proposes two ways of using the model transformations. One consists of optimizing a given system abstract architecture according to expressed developers' requirements in terms of QoS. The second consists of adapting an architecture according to a Grid middleware. Respectively:

- Each QoS property is represented by a design pattern. This representation is then adapted to the current software architecture by refinement.
- Each platform is represented by a design pattern and corresponding architectural properties. The system software architecture is then adapted to this platform by refinement as well.

To do so, the ARL expressiveness had to be extended with respect to the Grid domain. The next sections details our complementary semantic and its usage.

### 4.4. Grid Domain Specific Language

Enabling the *gMDE* requires the expression and consideration of new semantics. Indeed, as mentioned in section 3.1, the "Platform Independent View" and "Constraint View" are based on different meta-models. Thus the *gMDE* approach uses a Domain Specific Language (DSL) based on the SOA paradigm, which is a specialisation of the ARL language.

Figure 3 shows the different meta-models and their mapping, allowing the description of proper Grid services and their associated constraints. As a consequence the system architecture (*GEIM*) is respectively considered as a set of services by the software architect and is then mapped to the component-connector representation, which is computable. This paradigm mapping is a key element in our *gMDE* approach. The software architect can focus mainly on his domain requirements and benefits of the architecture-centric facilities to refine his system.

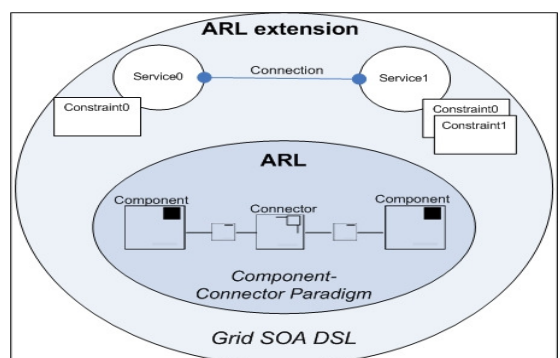

Figure 3: The grid domain specific language

As an example, the following is a generic description of a Grid architecture (voluntary simplified):

```
gridArchitectureRef is GridSOAArchitecture where {
  structure is {
    serviceName is style serviceTypeRef where {
      structure is { … service internal structure description … }
      connection is { … service connections descriptions … }
      constraint is { … QoS and / or platform constraints mappings … }
    } …
  }
  link is {
    attach serviceName0 to serviceName1 .
}}
```

The Grid architecture hereinbefore referred to as *gridArchitectureRef* is expressed in terms of services (e.g. referred to as *serviceName*), structure, connections and constraints.

Like the *GEIM* model, the *GECM* and *GETM* models are expressed using the same semantic. Following is the meta-model representing a constraint (of type QoS or Grid middleware).

```
constraintName is constraintTypeRef {
    on a:architecture actions {
        actionRef elemRef is typeRef {… element description …    }
    on b:architecturalElement actions {
        actionRef b .
        actionRef b .
    …}}…
```

The constraint, referred to as *constraintName,* is specified in terms of architectural elements (e.g. referred to as *elemRef)* providing its core functionalities and high-level refinement actions (e.g. referred to as *actionRef*) to be applied to one or more target elements "*b*". Using this semantic, a wide variety of QoS and Grid platforms constraints can be expressed and concretely used along the *gMDE* design process. Given the flexibility of our formal model-driven approach and relying on the correctness of our models, the resulting technique is able to tackle every aspect of software architecture transformations needed in Grid developments. These models and the enacted gMDE design process constitute the core of our *gMDE* environment (called *gMDEnv* – not detailed in this paper).

## 5. THE MDEGRID EXAMPLE

In order to demonstrate the core gMDE concepts, we introduce here the mdeGrid system example. For clarification, this example only treats the application of one QoS constraint model.

The *mdeGrid* system aims at providing clients, round-the-clock access to data stored in the Grid.

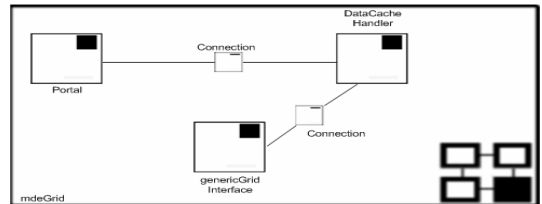

Figure 4. The mdeGrid system architecture

To provide such functionality, the *mdeGrid* system software architecture has been described as a set of services with dedicated roles. As detailed in the figure 4, the system architecture features three main services:
- The *Portal* : in charge of delivering data and answering to clients' requests. This service handles interactions with other Grid services in order to satisfy the client's request.
- The *DataCacheHandler* : this service collects and caches data queried from the grid through the *genericGridInterface* service. It updates this data automatically by checking it periodically and downloading if necessary.
- The *genericGridInterface* : this service represents the interface to a given grid middleware. (*NB:* this interface is considered as generic until the Grid platform is selected as explained later in the example).

```
archetype mdeGrid is architecture {
  types is {…}
  ports is {…}
  behaviour is {
    archetype Portal is component {…} .
    archetype genericGridInterface is component{…} .
    archetype DataCacheHandler is component {
      types is { type Data is any . type resultSet is tuple [String, String] }
      ports is {
        archetype ComsP0 is port {
          incoming is {ComsIncP0C0 is connection ( resultSet )}
          outgoing is {ComsOutP0C0 is connection ( Data )}
        } .
        archetype ComsP1 is port {
          incoming is {ComsIncC0 is connection (Data)}
          outgoing is {ComsOutC0 is connection (resultSet)}
        } }
      behaviour is {
        --<faulttolerance::priority:1,range:1>--
        value resultSet is connection (Data);
          value query := "the query expression…";
          recursive value readGridDBEntries is abstraction();
          {
            via ComsOutP0C0 send query;
            via ComsIncP0C0 receive res:resultSet;
            updateLocalCachedDB(res);
            readGridDB();
          };
          recursive value clientDataRequest is abstraction();
          {
            via ComsOutP1C0 receive clientRequest:request;
            res := processClientRequest(clientRequest);
            via ComsIncP1C0 send res;
            cacheClientResultSet(res);
            clientDataRequest();
          }; …
```

```
        compose {
            readGridDB() and clientDataRequest()
        }}}}
    unifies DataCacheHandler::ComsP1::ComsIncC1
        with Portal::PortComsP0::PortComsOutC0 ... }}
```

Figure 5. The mdeGrid architecture specification

Using our DSL, the system software architecture is specified and then transformed into ARL (see figure 5), which constitutes the *GEIM* model - presented in section 3.1. (*NB*: for clarification, the software architecture description is simplified, i.e. types, ports, connections and behaviours are not expressed). Once the system architecture is specified, the software architect can express non-functional requirements. As an instance, it is relevant in the *mdeGrid* architecture to ensure fault-tolerance over the *DataCacheHandler* service to guarantee uninterrupted data access to clients. To explicitly indicate that this service should be fault-tolerant, the software architect assigns it a constraint mapping. The mapping declaration is split into three parts as detailed below. The first part specifies its nature, the second its priority with respect to other constraints and thirdly its range/level. This constraint mapping is attached to the architectural element as an annotation inside the *GEIM* model following this scheme: "--<constraintRef::priority:#,range:#>--" (see the *DataCacheHandler* architectural element description in figure 5 for a detailed example of mapping). Once analyzed during the *gMDE* design process, the corresponding constraint design pattern is selected by the system (see figure 6). The following definition is a simplified representation of the Fault-Tolerance design pattern:

```
FT is qualityOfServiceProperty {
    on mdeGrid:architecture actions {
        include FTConnector is connector {
            ... connector architectural description ...}
        on DataCacheHandler :architecturalElement actions{
            replicate DataCacheHandler to DataCacheHandlerClone0;
            unify DataCacheHandler::ComsP0::ComsOutC0 with
                FTConnector:: genericGridInterfaceComsP0::genericGridInterfaceIncC0 .
        }}...
```

Figure 6. The fault-tolerance GECM

Our engineering environment (*gMDEnv*) then proceeds to the elaboration of the transformation model needed to fix the non-functional requirement. This is what is shown in figure 7. Inside the *gMDEnv*, a model-driven approach is enacted for the predictive non-functional and functional analysis of architectural elements. From the original *GEIM* model, the system analyzes the different constraint mappings and generates a satisfactory model of atomic transformations to apply with respect to the corresponding constraint design pattern.

The analysis conducted by the system is an heuristic method to determine constraint compatibilities and solutions among architectural elements and design patterns. The system tries to map constraints between architectural elements through inference rules and selects which transformation is the best suited. This iterative process leads progressively to the elaboration of a satisfactory transformation model applicable in context. This transformation is explained in figure 7 and constitutes an example of the first part of the *gMDE* design process. In the resulting architecture (the *GEIM'*), the fault-tolerance has been provided by the introduction of a new connector "*FTConnector*" – a representation of a known pattern for fault-tolerance handling - and the replication of the *DataCacheHandler* architectural element as a recovery service.

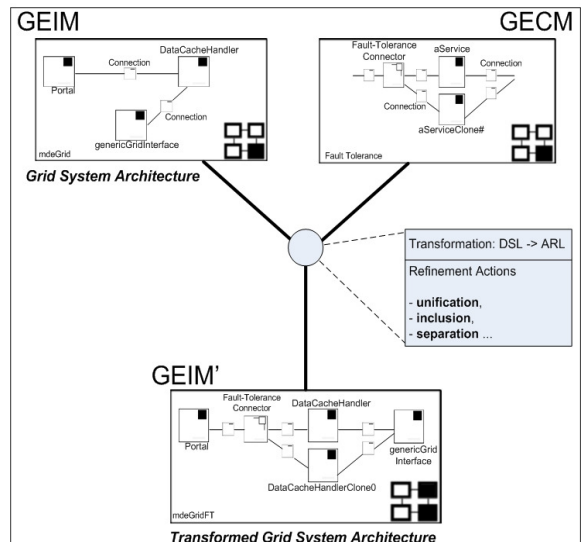

Figure 7. A gMDE model transformation

Figure 8 details the obtained new *mdeGrid* system description:

```
behaviour is {
    archetype Portal is component {...} .
    archetype genericGridInterface is component{...} .
    archetype DataCacheHandlerClone0 is component {...} .
    archetype DataCacheHandler is component {
        behaviour is {
```

```
        archetype FTConnector is connector { …
          behaviour is {
            recursive value availabilityChecking is abstraction();
            {
                if (serviceDown) value serviceRedirectionURL :=
                              DataCacheHandlerClone0;
                              availabilityChecking();
                };
                compose { availabilityChecking() }
              } .
            recursive value readGridDBEntries is abstraction();
            {…};
            recursive value clientDataRequest is abstraction();
            {…}; …
              compose {readGridDB() and clientDataRequest()}… }
```
Figure 8. The DataCacheHandler new behaviour

Thus, the clients' requests (through the *Portal* service) are re-directed to the clone service in case of a service failure. The same approach is undertaken when adapting the specified system architecture to a particular Grid middleware. The *genericGridInterface* architectural element is refined by model transformation so that the system architecture satisfies the architectural constraints implied by the design pattern.

As a conclusion, the model-driven paradigm enables the introduction of well-known design patterns for every aspect whether functional or not. For example, other patterns can be introduced for non-functional requirements like load balancing, security, performance, cost policies etc. However, for simplification matters, we do not discuss these in this paper although they are treated in our *gMDE* engineering environment (*gMDEnv*).

## 6. OUTLOOK AND CONCLUSION

In this paper we presented a technique for specifying Grid applications by modeling and by transforming these models to automate their adaptation to specific platforms and QoS constraints. We introduced an example to illustrate how the approach tackles QoS specifications in addition to platform requirements. Our investigation has lead to the elaboration of a wide range of frequently used Grid platforms and QoS constraint models. The efficiency of the approach relies strongly on the correctness of these models; consequently great care is being taken to ensure this. As a proof of concept, the engineering framework being developed (*gMDEnv*) enacts the combination of the formal architecture-centric and model-driven approaches introduced previously. In its current state, it is already capable of handling most of the presented models and transformations.

Since this approach is based on the concepts of re-use and execution platform independence, our engineering framework scope is not limited to the Grid domain. The same approach can tackle other developments based on the SOA vision such as web service-based applications (i.e. online traders, booking systems, video on demand systems etc). Thus, the benefits of using the *gMDE* are numerous. Formal application models designed using our framework are persistent and re-usable. For instance, one can use libraries and previously stored models to design new applications. The approach is scalable; one can extend the scope limitation of the framework by providing the corresponding new constraint and mapping models. From the establishment of well-known architectural concepts, the framework brings a high level of description to the user while promoting user-friendliness through a simple semi-automated graphical user interface (see figure 9).

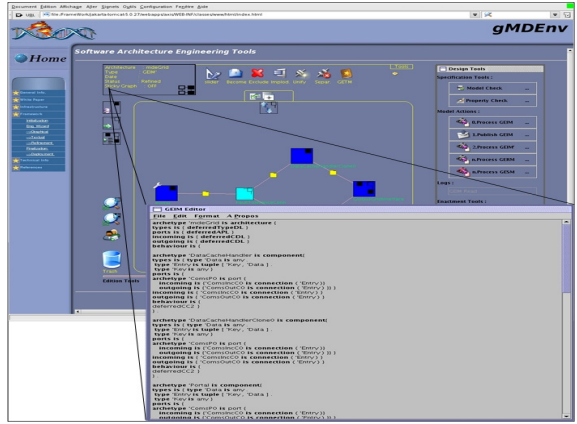

Figure 9. The gMDEnv graphical user interfaces

Finally, with respect to model transformations, an interesting area of future research is the development of a decision system to support users through model-driven transformations. Indeed, some of the adaptations required to satisfy platforms and QoS constraints can lead to critical decisions. We are using examples such as the one described in section 5 and the MammoGrid development experience (Amendolia et al, 2005), to elicit the framework requirements. The *gMDEnv* and the presented approach are currently in use to evaluate potential advantages in the development process of the MammoGrid application. There are clearly identified issues in the development of MammoGrid on which the *gMDEnv* emphasizes, such as adapting the system to other Grid platforms, improving the global application security level or porting the system to different programming languages. From these case studies, the preliminary conclusions are encouraging and show the relevance of this formal model-driven paradigm applied to the Grid domain.

This paper is a first investigation of the model-driven paradigm enactment using established formal architecture-centric concepts. Besides supporting the usefulness of the ArchWare ARL language, we are able to draw a number of conclusions. We learned that the model-driven approach is a very useful paradigm when addressing cross-platform developments and problems of re-use but it must be dependent on a rigorous basis to be efficient. The formal dimension brought by ArchWare is one of the key points of our successful implementation, especially in using a formal refinement language. Similarly we learned that QoS attributes are not easy to quantify in models. There is a true lack of standards that could help significantly when considering resource comparisons. In the context of other engineering frameworks and given the concepts we have now in hand, our approach can provide relevant benefits to the practice of Grid system engineering. From our experience, we believe that the presented approach is an important contribution to the development of new Grid systems.

## ACKNOWLEDGMENTS

The authors wish to thank their Home Institutions and the European Commission for financial support in the current research.